\documentstyle[psfig]{mn}

\newif\ifAMStwofonts

\newcommand{\lapp}{\mbox{\raisebox{-0.3em}{$\stackrel{\textstyle <}{\sim}$}}}
\newcommand{\gapp}{\mbox{\raisebox{-0.3em}{$\stackrel{\textstyle >}{\sim}$}}}

\title{A radio study of the superwind galaxy NGC~1482}
\author[Ananda Hota and D.J. Saikia]
       {Ananda Hota$^1$,$^2$\thanks{hota@ncra.tifr.res.in} and D.J. Saikia$^2$\thanks{djs@ncra.tifr.res.in} \\
$^1$ Joint Astronomy Programme, Indian Institute of Science, Bangalore 560 012, India \\
$^2$ National Centre for Radio Astrophysics, TIFR, Pune University Campus, Post Bag 3, 
Pune 411 007, India}
\date{Accepted.    Received }

\pagerange{\pageref{firstpage}--\pageref{lastpage}}
\pubyear{  }

\begin{document}

\maketitle

\label{firstpage}

\begin{abstract}
We present multifrequency radio continuum 
as well as \hbox{H\,{\sc i}} observations of the
superwind galaxy NGC1482, with both the Giant Metrewave Radio Telescope (GMRT) 
and the Very Large Array (VLA). This galaxy has a remarkable 
hourglass-shaped optical emission line outflow as well as bi-polar soft X-ray bubbles 
on opposite sides of the galactic disk. The low-frequency, lower-resolution radio 
observations show a smooth structure. From the non-thermal emission, we 
estimate the available energy in supernovae, and examine whether this would be adequate
to drive the observed superwind outflow. 
The high-frequency, high-resolution radio images of the central starburst 
region located at the base of the superwind bi-cone shows one prominent 
peak and more extended emission with substructure.  This 
image has been compared with the infrared, optical red-continuum,
H$\alpha$, and, soft and hard X-ray images from Chandra to understand 
the nature and relationship of the various features seen at different wavelengths.
The peak of infrared emission is the only feature which is coincident with the
prominent radio peak, and possibly defines the centre of the galaxy. 
 
The \hbox{H\,{\sc i}} observations with the GMRT show two blobs of  
emission on opposite sides of the central region. These are rotating about the centre of
the galaxy and are located at $\sim$2.4 kpc from it. 
In addition, these observations also reveal a multicomponent \hbox{H\,{\sc i}}-absorption
profile against the central region of the radio source, with a total width of 
$\sim$250 km s$^{-1}$. The extreme blue- and red-shifted absorption components are at 1688
and 1942 km s$^{-1}$ respectively, while the peak absorption is at 1836 km s$^{-1}$. This
is consistent with the heliocentric systemic velocity of 1850$\pm$20 km s$^{-1}$, estimated 
from a variety of observations.  We discuss possible implications of these results.
\end{abstract}


\begin{table*}
 \centering
 \begin{minipage}{140mm}
  \caption{Basic Data on NGC~1482.$^a$}
  \begin{tabular}{@{}cccccccc@{}}
\hline
 RA (J2000)$^b$
& DEC (J2000)$^b$ & Type$^c$ &
 a $\times$ b$^d$ & V$_{sys}$$^e$ & i$^f$ & log(L$_{FIR}$)$^g$ & D$^h$ \\
 (h m s) & ($^\circ$ $^{\prime}$ $^{\prime\prime}$) &
& ($^\prime$ $\times$
$^\prime$) & (km s$^{-1}$) & ($^\circ$) & (L$_\odot$) & (Mpc) \\
\hline
 03 54 38.92  & $-$20 30 07.6   & SA0/a &
2.5 $\times$ 1.4 & 1850$\pm$20 & 58 & 10.66 & 24.7 \\
\hline
\end{tabular}\hfill\break
$a$ Taken from the NASA Extragalactic Database (NED), unless stated otherwise.
\hfill\break
$b$ The position of radio peak from our high-resolution, VLA A-array, 8460 MHz image \hfill\break
$c$ Morphological type.\hfill\break
$d$ Optical major and minor axes.\hfill\break
$e$ Heliocentric optical systemic velocity (Veilleux \& Rupke, private communication). This is consistent with
\hbox{H\,{\sc i}} observations of Roth et al. (1991) and those presented in this paper. \hfill\break
$f$ Inclination angle, from Strickland et al. (2004).\hfill\break
$g$ Log of the far infra-red luminosity (Condon et al. 1990) using the revised distance. 
    \hfill\break
$h$ Distance estimated using the galaxy recessional velocity with respect to the cosmic
    microwave background radiation and H$_0$=71 km s$^{-1}$ Mpc$^{-1}$ (Spergel et al. 
    2003). For this distance 1$^{\prime\prime}$=120 pc.
\end{minipage}
\end{table*}

\section{Introduction}
Galactic-scale outflows along the minor-axes of nearby edge-on
galaxies have been observed over a wide frequency range extending
from radio continuum to X-ray wavelengths. The more prominent outflows 
with an outflow mass and kinetic energy of approximately 
$10^{5} - 10^{7}$ M$_{\odot}$ and $10^{53} - 10^{55}$ ergs
respectively are often referred to as superwinds, and these have been observed
over a wide range of redshifts (e.g. Heckman, Armus \& Miley 1990; 
Pettini et al. 2001; Veilleux 2002a, 2003).
The  superwind could contain different phases of the metal enriched interstellar
medium (ISM) such as hot X-ray emitting gas, warm ionized gas emitting 
ultraviolet (UV) and optical emission lines, synchrotron emitting
relativistic plasma and magnetic fields, dust, as well as 
\hbox{H\,{\sc i}} and molecular gas  (Dahlem 1997; Strickland 2001). These
superwinds play an important role in heating and supplying
kinetic energy to the Inter Galactic Medium (IGM) and enriching it with metals.
Many  of these outflows are interpreted as the combined effect of
numerous supernovae and stellar winds in the nuclear starforming
region of the parent Seyfert or starburst galaxy. 
An active galactic nucleus (AGN) or winds from the tori in the nuclear regions
could also contribute
to the observed outflows (Heckman et al. 1990; Balsara \& Krolik 1993; Baum et al. 1993; 
Krolik \& Kriss 1995; Colbert et al. 1996a,b; Weaver 2001; Veilleux 2002b).

       In this paper we present a detailed radio study of the interesting 
superwind galaxy NGC1482, which has received relatively little attention. The
basic properties of this galaxy are summarised in Table 1. In an emission-line survey of
early-type spirals Hameed \& Devereux (1999) noticed the presence of ``filaments 
and/or chimneys of ionized gas extending perpendicular to the disk''. 
Veilleux \& Rupke (2002) imaged the galaxy in H$\alpha$ and \hbox{N\,{\sc ii}}
and highlighted the remarkable hourglass-shaped optical emission 
line outflow with a velocity of $\sim$250 km s$^{-1}$. They estimated the
energy in the optical emission-line outflow to be at least 2$\times$$10^{53}$ ergs. The 
ionization ratios of the gas in the superwind have been interpreted to be
due to shock formation in the outflow. More recently, Veilleux et al. (2003 and
private communication) 
have reported an outflow velocity of $\sim$460 km s$^{-1}$ in one of the filaments.
The soft X-ray image from the 
Chandra Observatory also exhibits bipolar emission along a similar axis to that of
the optical hourglass-shaped structure (Strickland et al. 2004).

       We have studied this galaxy at radio continuum wavelengths ranging
from 335 MHz to 14965 MHz as well as in \hbox{H\,{\sc i}} using both the GMRT and the
VLA. While VLA images at 1.49 GHz have been presented by Condon et al. (1990), 
continuum observations at lower and higher frequencies as well as \hbox{H\,{\sc i}}
images have been presented here for the first time.
The objectives of the observations were to clarify the radio structure, constrain
the energetics and explore evidence of outflow at radio wavelengths. We first briefly 
describe the observations in Section 2. The large-scale
radio structure, its comparison with the images at other wavelengths, and 
some of the implications are discussed in Section 3. The small-scale radio
structure and a comparative study of this with features seen at other 
wavelengths are also presented in Section 3. Section 4 describes the \hbox{H\,{\sc i}}
observations, where we report the detection of both emission and absorption
lines, discuss these results and compare with the spectral information at
optical and CO wavelengths. In Section 5 we summarise the conclusions.

\section{Observations and Data analyses}
The observing log for both the GMRT and VLA observations is presented in Table 2,
which is arranged as follows. Column 1: Name of the telescope where we also list
the configuration for the VLA observations. Column 2: The frequency of the observations
where \hbox{H\,{\sc i}} denotes spectral line observations centred at an 
observed frequency of $\sim$1412 MHz. Columns 3 and 4: Dates of the observations and
the time, t, spent on the source. Columns 5 and 6: The phase calibrator used
and its flux density. A conservative estimate of the error in the flux density is 
approximately 10\% at 335 and and 5\% at the higher frequencies.   

\begin{table}
  \caption{Log of the radio observations}     
  \begin{tabular}{l r c r c c}
\hline
 Telescope    & Freq.             & Obs.       &      t         & Phase           & S$_{\rm cal.}$   \\
              & MHz               & date       &    min         & Calib.          &            Jy   \\
\hline
GMRT          &335                & 2002Aug17  &    60          &0521$-$207         &  9.67           \\
GMRT          &615                & 2003Mar07  &    180         &0453$-$281         &  2.21           \\
VLA-BnA       &1365               & 2002May20  &    25          & 0416$-$188        &   2.85          \\
GMRT      &1409,\hbox{H\,{\sc i}} & 2002Nov16  &    180         & 0453$-$281        &   1.99          \\
VLA-A         &4860               & 2003Jun26  &    54          & 0416$-$188        &   0.68          \\
VLA-BnA       &8460               & 2002May20  &    23          & 0416$-$188        &   0.68          \\
VLA-A         &8460               & 2003Jun26  &    60          &0416$-$188         &   0.83          \\
VLA-A         &14965              & 2003Jun26  &    20          &0348$-$278         &   1.02          \\
\hline
\end{tabular}
\end{table}

\begin{figure*}
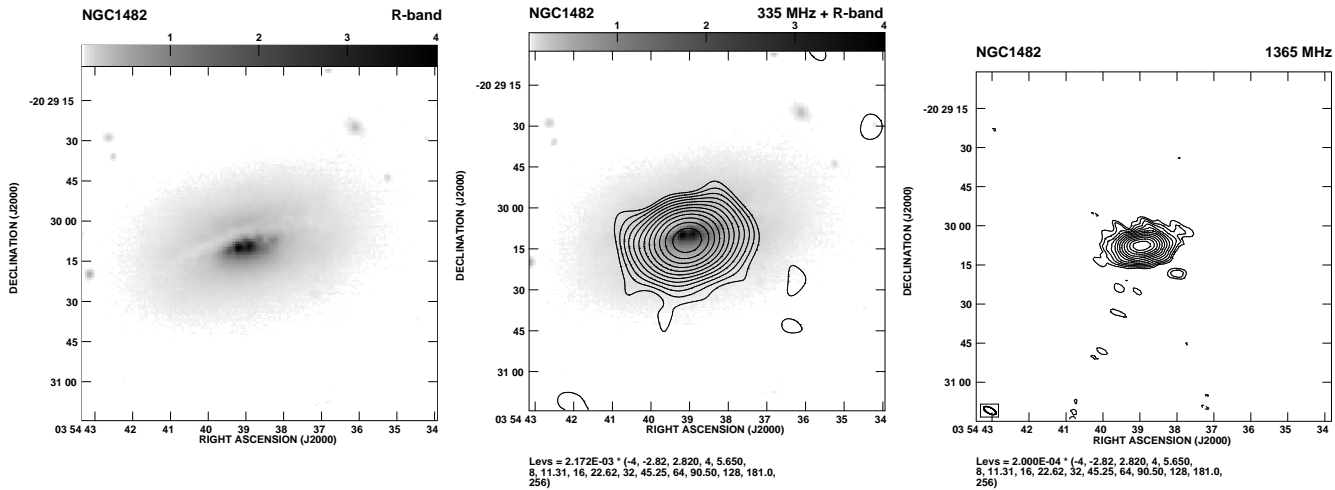

\hbox{
  \psfig{file=NGC1482.R.SQ.PX.335.vlaLsize.PS,width=2.3in,angle=0}
  \psfig{file=NGC1482R.335.PX.SIZE.PS,width=2.3in,angle=0}
  \psfig{file=NGC1482.VLA.R335SIZE.PS,width=2.3in,angle=0}
   }
\caption[]{Left panel: An optical R-band image of NGC1482 .
           Middle panel: The GMRT  335-MHz contour image restored with a circular
beam of 11.6 arcsec superimposed on the optical
                         R-band image in grey scale.
           Right panel: The VLA BnA-array image at 1365 MHz with an angular resolution
of 5.13$\times$2.23 arcsec along a PA$\sim$62$^\circ$.
}
\end{figure*}

\subsection{GMRT}
The GMRT consists of thirty antennas, each of 45 m diameter, in an approximate `Y' shape
similar to the VLA but with each antenna in a fixed
position.  Twelve antennas are randomly placed within a central 1 km
by 1 km square (the ``Central Square'') and the remainder form the
irregularly shaped Y (6 on each arm) over a total extent of about 25
km.  Further details about the array can be found at the GMRT website
at {\tt http://www.gmrt.ncra.tifr.res.in}.  The observations were made
in the standard fashion, with each source observation interspersed
with observations of the phase calibrator.  The primary flux
density calibrator was 3C286 whose flux density was estimated on the
Baars et al. (1977) scale, using the latest (1999.2) VLA values. 
The bandwidth of the continuum observations at 335 and 615 MHz was 16 MHz,
while for the L-band observations it was 8 MHz.
The data analyses were done using the
Astronomical Image Processing System (AIPS) of the National Radio
Astronomy Observatory. Since GMRT data is acquired in the spectral-line mode,
gain and bandpass solutions were applied to each channel before combining them. 
The 335 MHz observations were affected by ionospheric
phase fluctuations. A significant amount of data had to be edited, and the
first round of phase self-calibration was done assuming a point source model.
In the subsequent runs of self-calibration the image of the target source
was used as the model. The position of the source was determined by aligning
its peak with that of the 615 MHz image. 

The analyses of the \hbox{H\,{\sc i}} observations
were also done in the standard way. 3C286 was the primary flux density and
bandpass calibrator. The total bandwidth for \hbox{H\,{\sc i}} observations
was 8 MHz and the spectral resolution was 62.5 kHz, which corresponds to
13.3 km s$^{-1}$ in the centre of the band.  We discarded any antennas with more than 3$\%$
fluctuations in the bandpass gains during the observations. A few channels in the
beginning and approximately ten channels towards
the end were also not included in the analyses. The AIPS task UVLIN was used
for continuum subtraction and the multi-channel data were then CLEANed using IMAGR.

Some of the observed parameters of the GMRT and the VLA continuum images are presented
in Table 3 which is arranged as follows. Columns 1 and 2 are similar to that
of Table 2, except that we also list the parameters from the NRAO VLA
Sky Survey (NVSS). Columns 3 to 5: The resolution of the image with the major and
minor axes being listed in arcsec and the position angle (PA) in degrees.
Column 6: The rms noise in units of mJy/beam. 
Columns 7 and 8: The peak and
total flux densities in units of mJy/beam and mJy respectively. In addition to the
rms noise in the image, we also examined the change in flux density by varying the
size of the box around the source. The error in the 
flux density is approximately 10\% at 335 MHz and 5\% at the higher frequencies. 

\subsection{VLA}
The VLA observations were also made in the standard way with a phase calibrator
being observed before and after each scan on a source. The primary flux density
calibrator was 3C286. However, since 3C286 is significantly resolved on the longer
baselines at 14965, 8460 and 4860 MHz with the VLA A-array, the following procedure
was adopted. The flux density of 3C286 was calculated by the AIPS task SETJY. The
flux density of the phase calibrator was estimated by comparing its visibility
amplitude with those of 3C286 at the shorter baselines, over the {\it uv} ranges
specified in the VLA Calibrator manual.  
These values of the phase calibrator flux density
were incorporated using the task GETJY, and are also listed in Table 2. We attempted
to make self-calibrated images for all the different data sets. However, the A-array
images did not improve after self calibration. Therefore, BnA-array images presented
here are self-calibrated ones, while the A-array ones are not.

\section{Radio continuum emission}
\subsection{Large-scale structure}
In Fig. 1 we show the optical R-band image of the galaxy (Strickland et al. 2004;
and Strickland, private communication),
our GMRT 335-MHz contour map superimposed on the optical image and our VLA BnA-array 1365-MHz 
image. The GMRT 335-MHz image, which has an angular resolution of 11.6 arcsec shows the source to be resolved  
with a deconvolved angular size of 14$\times$8 arcsec along a PA of 107$^\circ$. Its angular dimensions 
are less than
that of the optical galaxy, and shows no evidence of a large-scale outflow or halo due to diffusion of
cosmic ray particles. This is consistent with the NVSS image where the source appears to be a single
source with a 45 arcsec beam. The GMRT 615-MHz image is essentially similar to the 335-MHz image 
and is not shown here, although the flux density estimates are listed in Table 3. 
The somewhat higher-resolution VLA 1365-MHz image, which has a lower rms noise than the low-frequency
GMRT images, shows the central region to be very clearly extended along a PA$\sim$105$^\circ$, 
but again with no evidence of any large-scale diffuse emission. 

A comparison of the radio image with the Chandra soft X-ray bubbles of plasma shows that the
X-ray emission extends far beyond the radio extent as seen in the 335-MHz image (Fig. 2). The 
non-thermal synchrotron emission appears largely confined to the disk of the galaxy, with no 
significant emission from the outflowing superwind plasma. 

The GMRT images do not appear to have missed significant emission from any diffuse component.
A comparison of the GMRT flux densities with the known and reliable flux density measurements
listed by NED (Fig. 3) shows that these values are consistent with the overall straight, 
non-thermal spectrum
of the source with a spectral index, $\alpha$, of 0.82 (S$\propto\nu^{-\alpha}$) between 335 
and 1400 MHz.  We have not attempted to make a spectral index image from 
the low-resolution images since the number of beamwidths along the source is small.

Adopting the spectral index of 0.82 as the mean value over the region of emission, and assuming
the synchrotron emission to have lower and higher cutoffs at 10$^7$ and 10$^{10}$ Hz respectively,
a proton to electron ratio of unity, a filling factor of unity and an oblate spheroidal distribution for
the emitting region we estimate the minimum energy density and equipartiton magnetic field to be 
$\sim$5.9$\times$10$^{-11}$ ergs cm$^{-3}$ and 25 $\mu$G respectively. The radiative life 
time of an electron 
radiating in this field at 1.4 GHz is $\sim$2.2$\times$10$^6$ yr. These values are similar to estimates for
nearby galaxies (e.g. Condon 1992). The proton to electron ratio is not well determined for
external galaxies. For a value of $\sim$50 from studies of cosmic rays in our own 
Galaxy (e.g Webber 1991), the equipartition magnetic field would increase by a factor
of $\sim$2.5.  

Since the overall spectral index is steep, the contribution of the thermal fraction to the
total emission is expected to be small. Using the expression due to Condon (1992) for the 
thermal fraction in spiral galaxies, as measured globally, we estimate that only 4, 7 and 11
per cent of the total emission could be due to thermal components at 335, 615 and 1400 MHz
respectively. From the non-thermal emission, we estimate the supernova rate in this galaxy to
be $\sim$0.14 yr$^{-1}$ (Condon 1992). 
The supernova rate appears similar to estimates in other starburst galaxies
such as $\sim$0.1 yr$^{-1}$ for M82 (e.g. Huang et al. 1994), $\lapp$0.1 to 0.3 for
NGC253 (Ulvestad \& Antonucci 1997), $\sim$0.1 yr$^{-1}$ for the clumpy irregular
starburst galaxy Mrk 325 (Condon \& Yin 1990) and also for the starburst galaxy NGC3448 of
the Arp 205 system (Noreau \& Kronberg 1987) and $\sim$0.07 yr$^{-1}$ for NGC6951
(Saikia et al. 2002). 

We can examine whether the energy the from the estimated supernova rate is adequate to drive 
the observed superwind outflow. 
Adopting a value of 10$^{51}$ ergs for the kinetic energy of a supernova, 
the available energy over the dynamical lifetime of the bubble, $\sim $$ 7.5 \times 10^{6}$ yr,
is $\sim10^{57}$ ergs. However, the energy which drives the outflow depends on
the heating efficiency of the supernovae, i.e. the fraction of supernova energy which is not
radiated away. In a classic paper, Larson (1974) estimates that approximately 10 per cent of
the explosion energy is effectively transmitted to the gas. There is a wide range of values 
in the literature, with many of the simulations of galactic winds assuming a heating efficiency
of 100 per cent. Melioli \& de Gouveia Dal Pino (2004) 	note that this is often not consistent
with the observations, and that the heating efficiency may be time dependent and sensitive to
the initial conditions of the system, and cannot be assumed to be 100 per cent. 
For our purposes
we assume an efficiency of 10 per cent, so that the available energy is $\sim10^{56}$ ergs.
Veilleux \& Rupke (2002) estimate the kinetic energy involved in the outflow of the warm
ionized gas to be
$\gapp$2$\times$10$^{53}$n$_{e,2}^{-1}$ ergs, where n$_{e,2}^{-1}$ is the number density
in units of 100 cm$^{-3}$. They estimate the number density of particles in the entrained 
material to be $\lapp$100 cm$^{-3}$ from their [\hbox{S\,{\sc ii}}]$\lambda$6731/6716 line ratios. 
The total kinetic energy involved in the outflow could be significantly larger. In addition 
to the optical line-emitting gas and the known X-ray plumes, there could also be cold gas 
entrained in the flow. For example, for a
sample of far-infrared bright, starburst galaxies, Heckman et al. (2000) estimate the kinetic
energy in the cool gas to be $\sim$10$^{55}$ ergs. In the case of NGC1482, we need a 
reliable estimate of total energy in the outflow  before determining whether the observed 
supernova rate is adequate to drive the outflow.

\begin{figure}
\hbox{
  \psfig{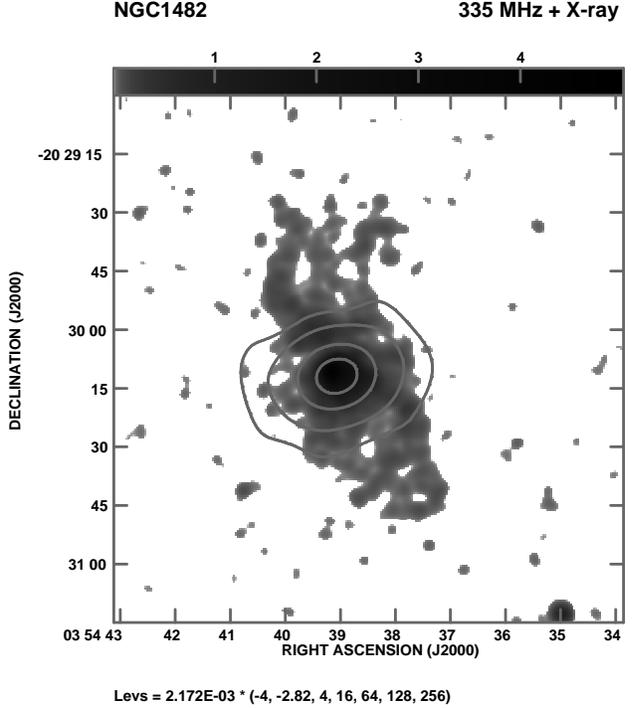}
   }
\caption[]{The GMRT 335-MHz contour image with an angular resolution of 11.6 arcsec superimposed on the soft 
X-ray image from Chandra X-ray observatory.}
\end{figure}

\begin{table}
  \caption{Observed parameters of radio continuum images}
  \begin{tabular}{lr rrr rrr}
\hline
 Telescope    & Freq.   & \multicolumn{3}{c}{Beam size}          &  rms  & S$_{\rm pk}$   & S$_{\rm tot.}$ \\
              & MHz     & maj.    & min.      & PA               &  mJy  &    mJy         &    mJy       \\
    &         & $^{\prime\prime}$ & $^{\prime\prime}$ & $^\circ$ &  /b   &   /b           &                \\
\hline
GMRT          &335      &11.6     &11.6       &  0            &2.24      &   370          & 723     \\
GMRT          &615      & 8.24    &5.44       &178            &0.65      &   140          & 397     \\
VLA-BnA       &1365     &5.13     &2.23       & 62            &0.17      &    35          & 198     \\
NVSS          &1400     &45.0     &45.0       &  0            &0.50      &    23          & 224     \\
GMRT          &1409     &2.96     &2.01       & 25            &0.60      &    21          & 218     \\
VLA-A         &4860     &0.64     &0.35       & 12            &0.04      &    1.9         & 44       \\
VLA-BnA       &8460     &1.06     &0.55       & 46            &0.07      &    1.3         & 25       \\
VLA-A         &8460     &0.36     &0.20       &  9            &0.02      &    0.9         & 18       \\
VLA-A         &14965    &0.20     &0.11       &  5            &0.09      & $<$0.3         &          \\
\hline
\end{tabular}\hfill\break
\end{table}

\begin{figure}
\hbox{
  \psfig{file=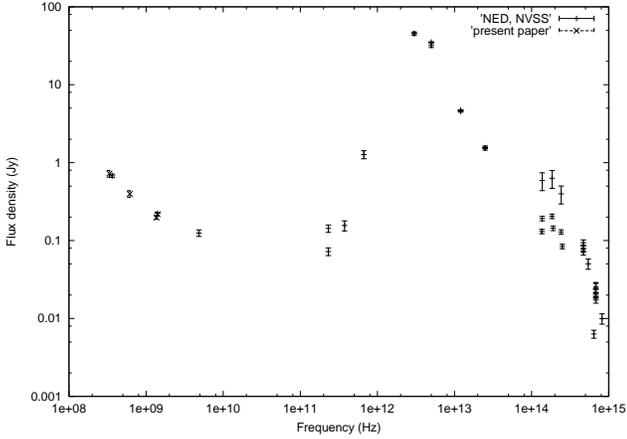,width=3.3in,angle=-90}
   }
\caption[]{The integrated spectrum of NGC1482 using values from NED, NVSS ($+$)and from the 
observations presented here ($\times$). 
           }
\end{figure}

\begin{figure}
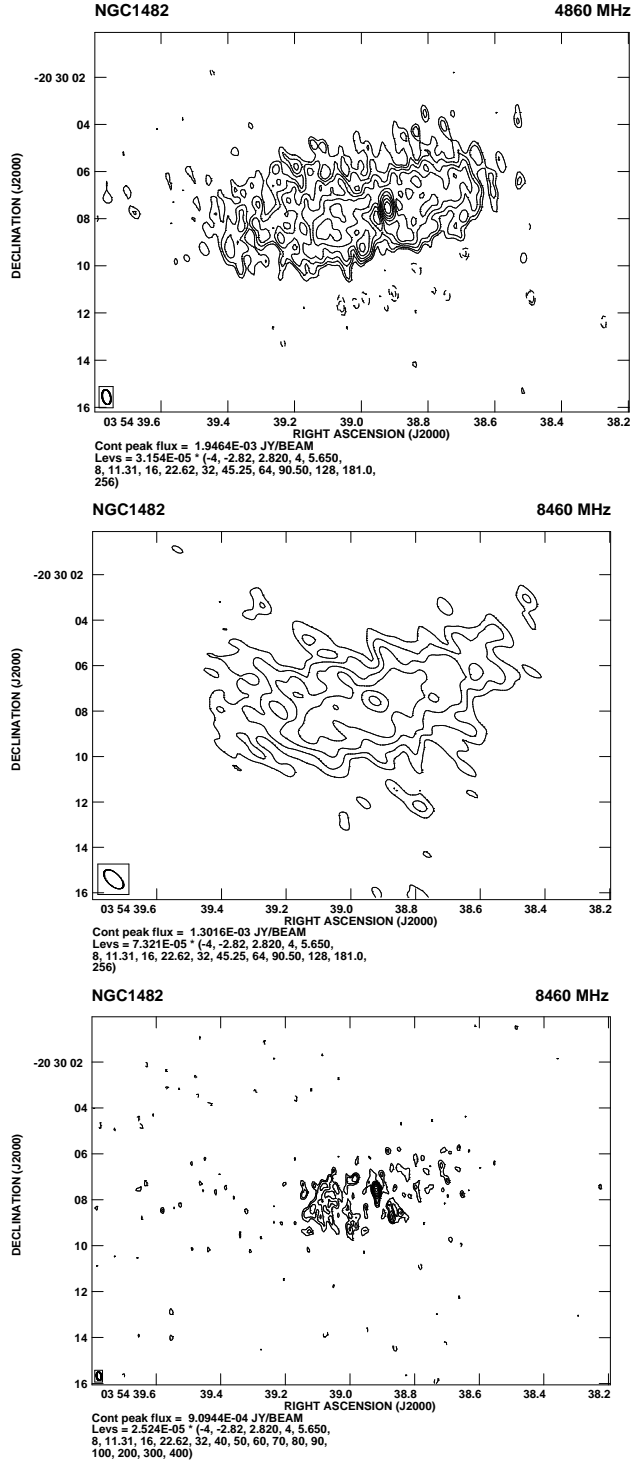
 
\vbox{
  \psfig{file=NGC1482VLAC.PS,width=3.3in,angle=-90}
  \psfig{file=NGC1482VLA.X.PS,width=3.2in,angle=-90}
  \psfig{file=NGC1482VLAXX.PS,width=3.2in,angle=-90}
   }
\caption[]{Top panel: VLA A-array image at 4860 MHz with an angular resolution of 0.64$\times$0.35 arcsec
                      along a PA$\sim$12$^\circ$. 
           Middle panel: VLA snapshot image with the BnA array at 8460 MHz. The angular resolution is
                      1.06$\times$0.55 arcsec along a PA$\sim$46$^\circ$.  
           Bottom panel: VLA A-array image at 8460 MHz with an angular resolution of
                      0.36$\times$0.20 arcsec along a PA$\sim$9$^\circ$.}
\end{figure}

\begin{figure} 
  \psfig{file=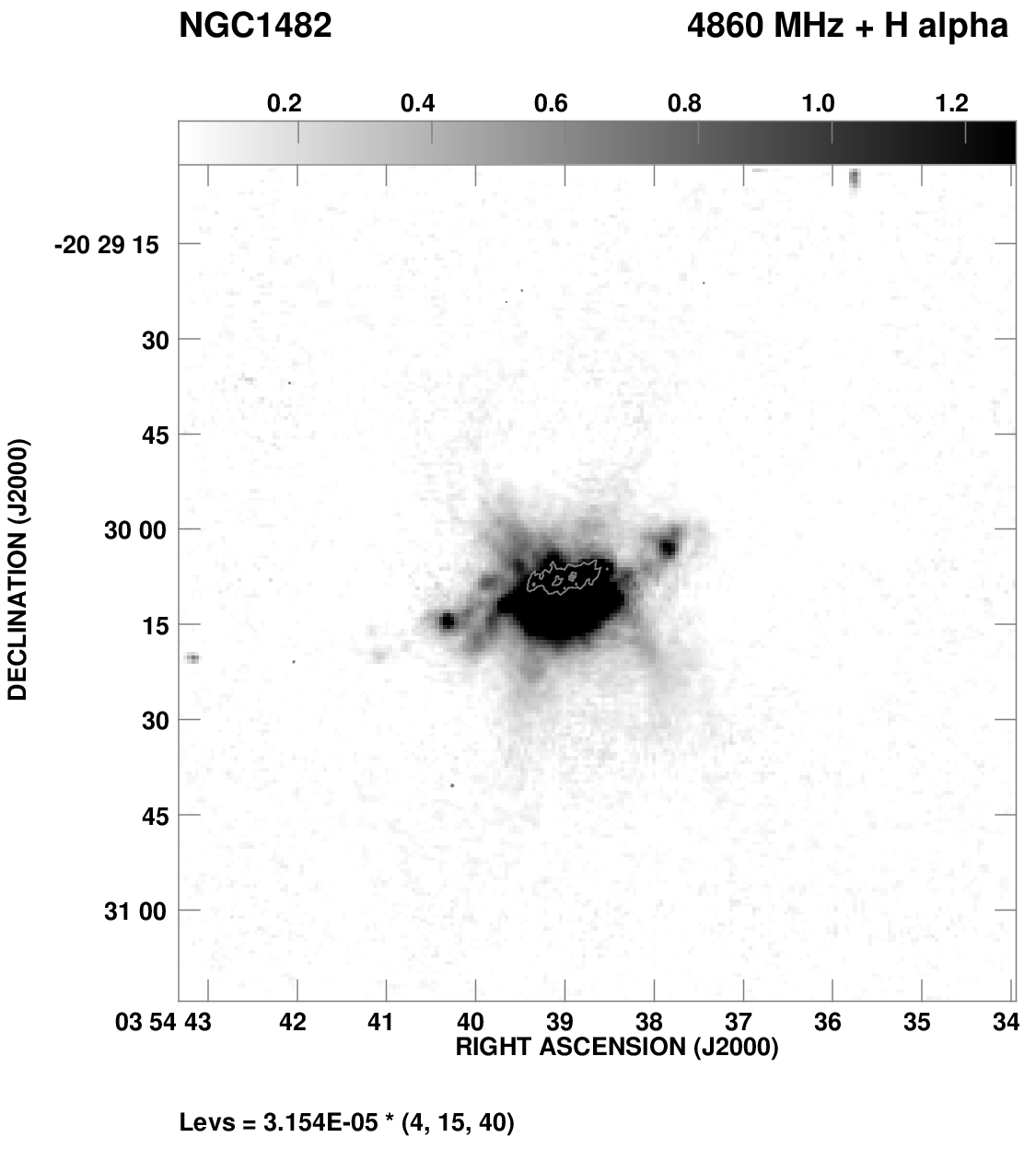,width=3.3in,angle=0}
\caption[]{The VLA A-array 4860 MHz image, superimposed on the 
grey scale image of H$\alpha$ emission. Only three of the radio contours are shown.}
\end{figure}

\subsection{Small-scale structure}
The 4860 MHz, VLA A-array image (Fig. 4) shows a prominent peak of emission with a brightness of 1.95
mJy/beam and more diffuse emission  along a PA of $\sim$100$^\circ$, similar to the
orientation of the galaxy. The diffuse emission is more prominent on the eastern
side of the peak, the total flux density on this side being 27 mJy
compared with 15 mJy on the western side. The brightest feature on the eastern side, which we
call the secondary peak, is located at RA: 03 54 39.12, Dec: $-$20 30 08.60 and has a 
peak brightness of 0.65 mJy/beam. 
This image shows a sharper boundary
on the southern side than on the northern one, as is also seen in the 
VLA BnA-array image at 1365 MHz (Fig. 1). The somewhat lower-resolution 
VLA BnA-array, 8460 MHz snap-shot image shows basically a similar 
structure with the peak brightness being 1.30 mJy/beam. Smoothing the 4860 MHz
image to that of the BnA array 8460 MHz one yields a peak brightness of 2.38  mJy/beam
and a spectral index of $\sim$1 for the prominent peak of emission. 
The spectral index of the central region between 4860 and 8460 MHz
is also $\sim$1, showing that the entire region has a steep radio spectrum. 

\begin{figure*}
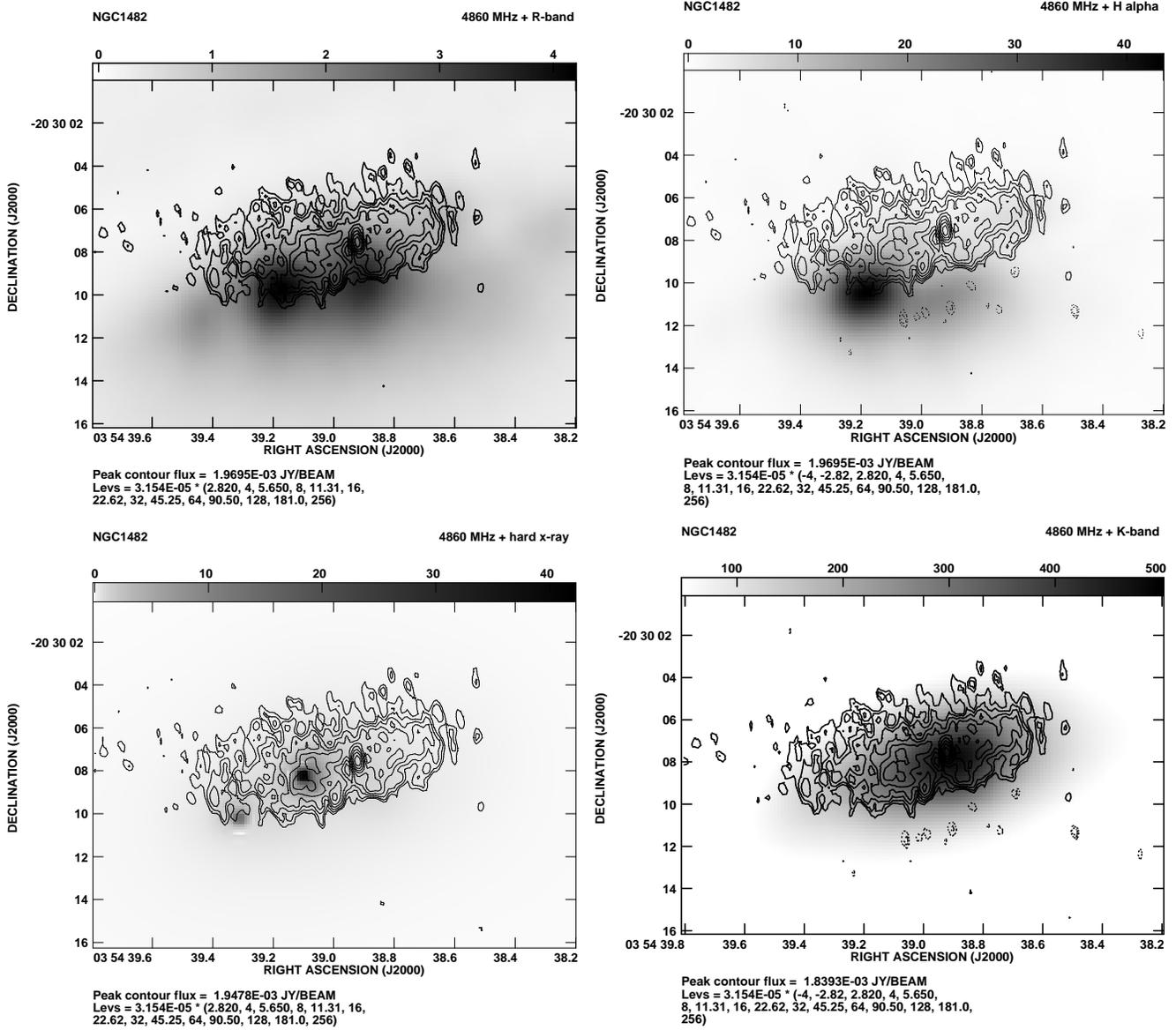

\vbox{
\hbox{
  \psfig{file=1482.R4X.C.NU.PS,width=3.45in,angle=-90}
  \psfig{file=1482HA4X.C.NU.X.PS,width=3.45in,angle=-90}
     }
\hbox{
  \psfig{file=1482.X4X.C.NU.PS,width=3.45in,angle=-90}
  \psfig{file=1482.K6X.C.NU.PS,width=3.45in,angle=-90}
     }
}
\caption[]{The 4860 MHz VLA A-array contour map with an angular resolution
           of 0.64$\times$0.35 arcsec along a PA$\sim$12$^\circ$ is shown superimposed on an
           optical R-band image (upper left), narrow-band H$\alpha$ image (upper right),
           the Chandra 2-8 keV hard X-ray image (lower left) and the
           infrared K-band 2MASS image (lower right) of only the central region of the galaxy.}
\end{figure*}

The expected brightness of this peak of emission at 14965 MHz with a similar  resolution 
to that of the VLA BnA-array image is $\sim$0.7  mJy/beam. Its  non-detection with 
a 3$\sigma$ upper limit of $\sim$0.3 
mJy/beam in the image with a resolution of $\sim$0.15 arcsec is consistent with this. 
Although its steep spectral index does not 
suggest it to be  an AGN, it is difficult to rule out the possibility of there 
being a weak AGN with more extended emission associated with it. It is relevant to note
here that from optical spectroscopic observations, Kewley et al. (2000) have classified it
to be a starburst galaxy without an AGN. The brightness temperature of the peak of emission,
estimated from the 4860 MHz image (e.g. Condon et al. 1991) is only $\sim$630$^\circ$ K 
making it difficult
to use it as a diagnostic to identify  whether it is an AGN. The VLA A-array 8460 MHz 
image with a resolution of $\sim$0.27 arcsec shows the peak feature to have a possible
extension to the south and again more emission on the eastern side of this peak (9 mJy)
compared with the western side (6 mJy). The peak brightness here is 0.9 mJy/beam.
The brightness temperature estimated for the peak is $\sim$740$^\circ$ K.
The linear resolution in the VLA A-array 8460 MHz image is approximately 30 pc, and its 
peak luminosity is  6.5$\times$10$^{19}$ W Hz$^{-1}$. The peak luminosity in the other regions of 
the source is typically 10$^{18}$ W Hz$^{-1}$. It is of interest to compare this with other 
galaxies. For example, the median luminosity at 8.4 GHz of the components observed in M82 which is
at a distance of 3.4 Mpc is $\sim$2$\times$10$^{18}$ W Hz$^{-1}$ 
(Huang et al. 1994; Allen \& Kronberg 1998) when observed with a linear 
resolution of $\sim$3 pc.
Considering a more distant galaxy, the median luminosity for the archetypal starburst
galaxy in the southern hemisphere, NGC1808 (distance$\sim$10.9 Mpc), is about
8$\times10^{18}$  W Hz$^{-1}$ with a linear resolution of 30 pc (Collison et al. 1994).
The luminosities in NGC1482 appear similar to the discrete features in some of 
the well-known archetypal starburst galaxies.

\subsection{Comparison with other wavebands}
The high-resolution radio images, which define the structure of the circumnuclear starburst
region, consist of one prominent peak of emission and several weaker secondary peaks.
A superposition of the VLA
4860 MHz image on the optical, hourglass-shaped structure traced by the  H$\alpha$ emission 
(Fig. 5) shows that the starburst region lies at the base of this structure, and
possibly provides the energy to drive the outflow.
The 4860 MHz image and the two H$\alpha$ peaks of emission at the extremeties appear to lie 
along a line perpendicular to the axis of the outflow. The 4860 MHz image appears displaced to the
north in the central visible H$\alpha$ region of emission; this apparent displacement 
could be largely due to obscuration of the northern part by the dust lane.

In order to understand the nature of the 
prominent radio peak we have compared the 4860 MHz radio image with the optical R-band continuum,
H$\alpha$, hard X-ray and infrared images (Fig. 6). The positions of the peaks in J2000 
co-ordinates at the different wavelengths are listed in Table 4. These positions have been
estimated from the FITS images kindly provided by Strickland and Veilleux (private communication). 
The R-band image shows two prominent peaks of emission with the peak brightness of the 
eastern one being brighter by a factor of $\sim$1.14, and a dust lane to the north. 
The northern part of the
the 4860 MHz image is seen through the dust lane. The R-band peaks possibly arise due to a combination
of enhanced emission in the sites of star formation and differential extinction in 
the circumnuclear region. These peaks
skirt the southern edge of the radio continuum image and are not coincident with the radio peak.
The H$\alpha$ image also shows two peaks of emission with  the maximum value for the eastern 
one being brighter than the western one by a factor of $\sim$2.
The H$\alpha$ peaks also lie along the southern boundary of the image, are within $\sim$1.$^{\prime\prime}$5
of the peaks in the R-band image, and are not coincident with the radio peaks. 

\begin{table}
\caption{Nuclear emission peaks at different wavebands}
\begin{tabular}{l c c }
\hline
Peaks & RA(J2000)  & Dec.(J2000) \\
\hline
Radio 4860 MHz & 03 54 38.92 & $-$20 30 07.5   \\
Radio 8460 MHz & 03 54 38.92 & $-$20 30 07.6   \\
X-ray West     & 03 54 39.10 & $-$20 30 08.2    \\
X-ray East     & 03 54 39.31 & $-$20 30 10.1     \\
H$\alpha$ West & 03 54 39.05 & $-$20 30 11.2   \\
H$\alpha$ East & 03 54 39.18 & $-$20 30 10.5  \\
R-band West    & 03 54 38.92 & $-$20 30 09.3    \\
R-band East    & 03 54 39.18 & $-$20 30 09.9    \\
\hline
\end{tabular}
\end{table}

The 2 to 8 keV hard X-ray Chandra image also shows two compact peaks of emission. Such
features have been seen in a number of nearby starburst galaxies such as M82 (Zezas et al. 2001), 
NGC4038/4039 (Zezas et al. 2002) and NGC3256 (Lira et al. 2002), and are 
interpeted to be due to X-ray binaries or supernova
remnants. In NGC1482, the western hard X-ray peak is within 0.5 arcsec of the secondary radio 
peak (see Fig. 6). The point source subtracted hard X-ray image (Strickland et al. 2004), not shown here, 
reveals more extended emission coincident with the secondary peak in the 4860 MHz image. These are possibly
due to one or more supernova remnants. The eastern hard
X-ray peak (see Fig. 6) lies at the southern edge of the radio image and does not have any compact component
associated with it. No hard X-ray peak is seen at the position of the prominent peak in the radio image.

A superposition of the radio image on the infrared 2MASS (2 Micron All Sky Survey) 
K-band image (Jarrett et al. 2003) 
shows the infrared peak
to be coincident with the prominent radio peak. The infrared image, which is largely due to a 
population of old giant stars, also appears asymmetric with more emission on the eastern side than
the western one. This asymmetry is similar to what is seen in the radio image, and reflects a greater
degree of circumnuclear activity on the eastern side of this peak. The coincidence of the radio and
infrared peaks suggest that this is possibly the centre of the galaxy. The centre of the galaxy
estimated from the outer isophotes of the R-band image is at RA 03 54 38.91 and Declination
$-$20 30 07.2, which is within an arcsec of the radio peak in the high-resolution images.
The geometrical centre of a line joining the peaks of the red- and blue-shifted 
\hbox{H\,{\sc i}} emission blobs  
(RA: 03 54 38.8 and Dec.: $-$20 30 10 in J2000 co-ordinates, Section 4.1) 
is within 2 arcsec of this position, and hence co-incident with it within the errors.

\section{\hbox{H\,{\sc i}} Observations}
The \hbox{H\,{\sc i}} spectra towards different regions of the galaxy 
show significant \hbox{H\,{\sc i}} emission from 
two diametrically opposite regions located at a distance of $\sim$20 
arcsec from the nuclear region, while \hbox{H\,{\sc i}} in absorption
is seen against the radio continuum source. These observations which have
an rms noise of $\sim$0.6 mJy/beam per channel are described below.

\subsection{\hbox{H\,{\sc i}} emission}  
The global \hbox{H\,{\sc i}} profile from a tapered image 
with an angular resolution of 8.19$\times$6.52 arcsec along
a PA of 36$^\circ$ is shown in Fig. 7. This spectrum has been obtained by 
specifying a box around the visible extent of the galaxy, and there is a hint
that it could be double-humped.  The 
heliocentric velocities at zero intensity range from $\sim$1690 to 2020 
km s$^{-1}$, indicating a systemic velocity of 1855$\pm$20 km s$^{-1}$.
The total width is $\sim$330 km s$^{-1}$, similar to that 
of typical spiral galaxies.
The published \hbox{H\,{\sc i}} spectrum by Roth, Mould \& Davies (1991) is very
similar and has a central velocity of 1859 km s$^{-1}$ with a velocity resolution
of 7.3 km s$^{-1}$. The total 
\hbox{H\,{\sc i}} flux density estimated from our spectrum is
9$\pm$1  Jy km s$^{-1}$, implying that the total \hbox{H\,{\sc i}} mass using the
standard expression (e.g. Mirabel \& Sanders 1988) is 
13.5$\pm$1.5$\times$10$^8$M$_\odot$. 

\begin{figure}
\hbox{
  \psfig{file=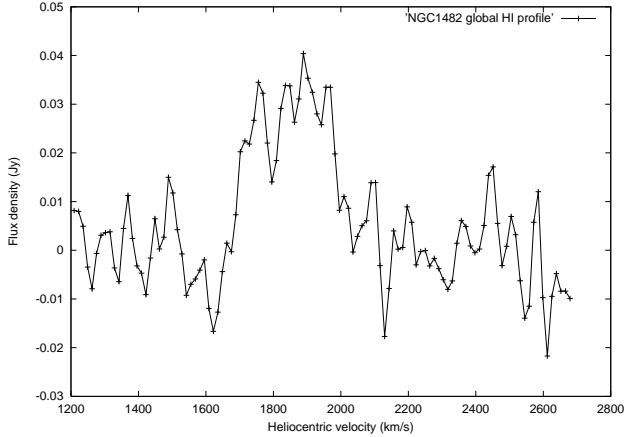,width=3.3in,angle=-90}
   }
\caption[]{The global \hbox{H\,{\sc i}} emission profile of NGC1482 obtained from a 
tapered image with an angular resolution of 8.19$\times$6.52 arcsec along a PA of 36$^\circ$.}
\end{figure}

We compare the global \hbox{H\,{\sc i}} profile with the CO J=1$-$0 spectra
published by a number of authors (Sanders et al. 1991; Elfhag et al. 1996;
Chini et al. 1996). Sanders et al. (1991) find a double-humped profile with the
blue-shifted component being narrower and weaker than the redshifted one, similar
to the \hbox{H\,{\sc i}} profile. The velocity width at the 20 per cent level 
of the peak is 325 km s$^{-1}$ . CO J=1$-$0 spectra published by Elfhag et al.(1996) and 
Chini et al. (1996) span a similar velocity range, but show significant
differences in their CO profiles. Chini et al. have noted the difference and have
mentioned that they do not have an explanation.

The estimates of heliocentric systemic velocity from \hbox{H\,{\sc i}} observations 
are consistent with those quoted from CO observations by Young et al. (1995) and
Elfhag et al. (1996). They find the values to be 1848 and 1847 km s${-1}$ respectively
with spectral resolutions of 10$-$15 km s$^{-1}$. These values agree with the estimate
of 1850$\pm$20 km s$^{-1}$ by Veilleux \& Rupke (private communication) from their
optical spectroscopic observatinos. The earlier estimate of 1916$\pm$39
km s$^{-1}$ obtained by fitting the H$\alpha$, H$\beta$, [\hbox{O\,{\sc iii}}], [\hbox{N\,{\sc ii}}]
and [\hbox{S\,{\sc ii}}] lines (Da Costa et al. 1991) appear to be higher. Da Costa et al.
note in their paper that Fairall has pointed out that 
some of their estimates differ from those in the literature by $\sim$100 km 
s$^{-1}$. At present, we adopt the value of 1850$\pm$20 km s$^{-1}$ as the heliocentric 
systemic velocity of the galaxy.

\begin{figure*}
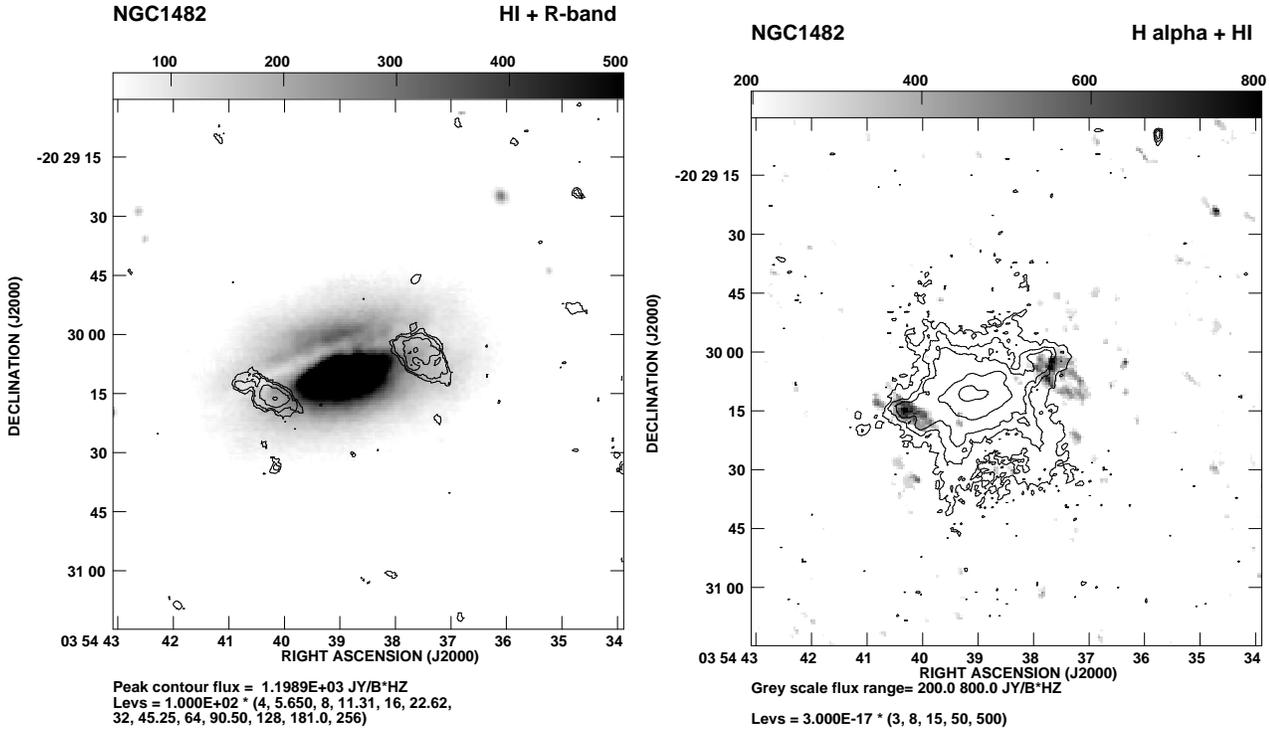

\hbox{
  \psfig{file=NGC1482.R.HI.X.SIZE.PS,width=3.3in,angle=0}
  \psfig{file=NGC1482.HA.HIT60.X.SIZE.PS,width=3.3in,angle=0}
     }
\caption[]{Left panel: \hbox{H\,{\sc i}} total-intensity contour map with an angular resolution 
of 8.19$\times$6.52 arcsec along PA of 36$^\circ$ superimposed on an optical R-band image.
Right panel: The \hbox{H\,{\sc i}} total-intensity image in grey with an angular resolution
of 6.95$\times$5.17 arcsec along PA of 34$^\circ$ superimposed on the 
H$\alpha$ contour map.}
\end{figure*}

The  \hbox{H\,{\sc i}} total intensity map has been generated from a MOMENT 
analysis in AIPS by integrating the velocities 
from 2090 to 1555 km s$^{-1}$ and blanking out points which are below three times
the rms noise.  A Hanning smoothing along the velocity axis and a Gaussian smoothing 
in the spatial plane were used for making this image. There are two blobs of 
emission on opposite sides of the nuclear region, and located about 20 arcsec
(2.4 kpc) from it (Fig. 8). 
van Driel (1987) and van Driel \& van Woerden (1991) have studied the distribution and
kinematics of a sample of lenticulars and early-type disk galaxies and compared their 
properties 
with late-type spirals. They note that while most barred spirals
have a pronounced central hole in their \hbox{H\,{\sc i}} distribution, often as large as the
bar itself, non-barred spirals and S0/a galaxies generally show filled \hbox{H\,{\sc i}} disks,
sometimes with a central depression. Although our results on \hbox{H\,{\sc i}} observations
of NGC1482, an S0/a galaxy, is broadly consistent with these trends, the depletion appears
pronounced, possibly due to the nuclear starburst.

At about 20\% of the peak intensity in the spectrum (Fig. 10), 
the estimated central velocity is 1835$\pm$20 km s$^{-1}$, consistent with the 
systemic velocity. The western 
blob has a maximum red shift of $\sim$190 km s$^{-1}$ relative to
the  systemic velocity of 1850 km s$^{-1}$, while the maximum blue-shift value for the 
eastern one is 
$\sim$220 km s$^{-1}$. The width of the line towards each blob is $\sim$250 km s$^{-1}$
at 20 \% of the peak intensity. 
The estimated column density and mass for the western blob are 
4.0$\pm$0.3$\times$10$^{21}$ atoms cm$^{-2}$
and 15$\pm$1$\times$10$^{7}$ M$_{\odot}$ respectively, while
the corresponding values for the eastern blob are 
3.8$\pm$0.3$\times$10$^{21}$ atoms cm$^{-2}$ and
13.5$\pm$1$\times$10$^{7}$ M$_{\odot}$ respectively.

The  \hbox{H\,{\sc i}} total-intensity distribution 
is shown superimposed on an optical R-band image of the 
galaxy as well as on the H$\alpha$ ionization cone in Fig. 8. 
It is clear that the  \hbox{H\,{\sc i}} blobs of emission lie at the
outer edges of the galaxy close to the edge of the dust lane. It avoids the central region 
where active star formation is taking place. 
A comparison of the \hbox{H\,{\sc i}} emission blobs with the H$\alpha$ image (Fig. 8)
shows that the \hbox{H\,{\sc i}} blobs define an axis which is approximately perpendicular
to the optical emission-line outflow bicone. The peaks of these two blobs are close to 
two H$\alpha$ knots or peaks of emission in the outer edges of the galaxy.  
The \hbox{N\,{\sc ii}} to  H$\alpha$ flux density ratios 
suggest that the two knots are consistent with these being \hbox{H\,{\sc ii}} regions
(Veilleux \& Rupke 2002). More sensitive \hbox{H\,{\sc i}} observations of higher angular
resolution are required to establish reliably the structure and dynamics of the blobs, 
and the effects of the H$\alpha$ knots and the circumnuclear startburst on these properties.

\subsection{\hbox{H\,{\sc i}} absorption}

\begin{figure*}
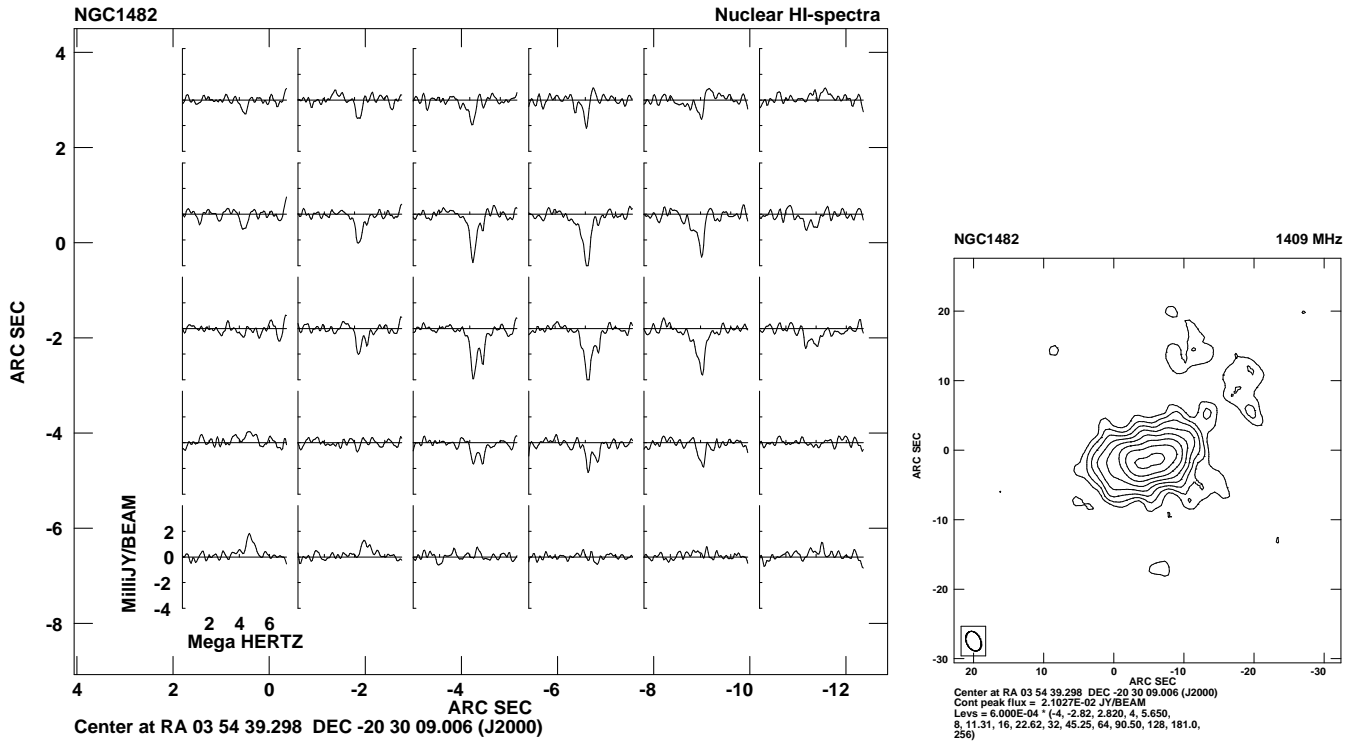

\hbox{
  \psfig{file=NGC1482.NU.PLCUB.PS,width=4.7in,angle=-90}
  \psfig{file=NGC1482.GMRT.3X2.PS,width=2.3in,angle=0}
     }
\caption[]{Left panel: Multiple \hbox{H\,{\sc i}} absorption spectra against the central 
radio continuum source with an angular resolution of 6.95$\times$5.17 along PA of 34$^\circ$.
Right panel: The GMRT 1409-MHz continum image against which \hbox{H\,{\sc i}}  absorption is seen.
This image has an angular resolution of 2.96$\times$2.01 arcsec along PA of 25$^\circ$.
}
\end{figure*}

In Fig. 9, we  present the absorption profiles towards different regions of the
continuum source with an angular resolution of 6.95$\times$5.17 arcsec along PA of 34$^\circ$,
and an rms noise of 0.6 mJy/beam.  A 21-cm continuum image with an angular resolution
of 3$\times$2 arcsec along a PA of 28$^\circ$ from the same data set is also shown.
The continuum image shows evidence of two peaks of emission
towards the nuclear region. The peak absorption occurs at a heliocentric velocity of 
1836$\pm$15 km s$^{-1}$, which is consistent with the heliocentric systemic velocity. 
An absorption feature at this velocity is seen over
most of the source. In addition there is a blue-shifted component
at a heliocentric velocity of 1688 km s$^{-1}$ which is prominent on the
eastern side. The western side shows a weaker red-shifted shoulder with a
heliocentric velocity of $\sim$1942 km s$^{-1}$. These features which are 
seen in the absorption spectrum against the nuclear region (see Fig. 10)
taken over a rectangular area of 7$\times$4 arcsec, highlights the possibility
that absorption is caused by multiple clouds moving with different velocities. 
Hydrodynamical simulations of starburst-driven superwinds show that the cool 
gas could expand laterally in the disk of the galaxy, be carried vertically 
outwards by the tenuous superwind or be entrained in the interface between
the hot, superwind fluid and the cool, dense ISM (e.g. Heckman et al. 1990, 2000).
For a systemic velocity of 1850$\pm$20 km s$^{-1}$, there is a suggestion of a mild
asymmetry in the absorption profile with the approaching component being blue shifted
by 162 km s$^{-1}$ while the receding component is red shifted by 92 km s$^{-1}$. In
addition to galactic rotation, the width of the absorption profile may also be 
due to cool, HI components which have been hydrodynamically affected by the superwind,
but this needs further investigation. 

The optical depth in
each velocity channel has been calculated using the background continuum flux density 
of 60 mJy from the area of 7$\times$4 arcsec. 
The peak optical depth estimated from this spectrum is 0.11. 
The total column
density is 2.8$\pm$0.1$\times$10$^{21}$ cm$^{-2}$, which  has been estimated in the ususal
way (e.g. Heckman, Balick \& Sullivan 1978) by integrating the
optical depth from 1675 to 1956 km s$^{-1}$, and assuming a spin temperature of 100$^\circ$ K. 
The corresponding total mass of the absorbing \hbox{H\,{\sc i}} clouds is 
6.3$\pm$0.3$\times$10$^6$M$_{\odot}$.   

\begin{figure}
\hbox{
  \psfig{file=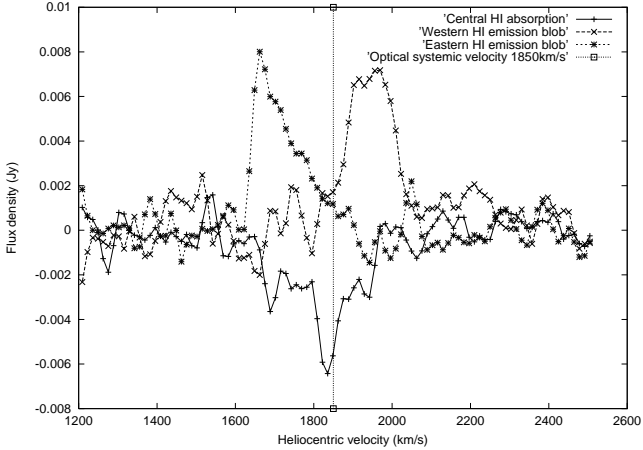,width=3.4in,angle=-90}
     }
\caption[]{The \hbox{H\,{\sc i}} emission  spectra of the western and eastern blobs, and
the \hbox{H\,{\sc i}} absorption spectrum against the central radio continuum source. 
The emission
spectra have been obtained from the image with an angular resolution of
8.19$\times$6.52 arcsec along PA of 36$^\circ$, while the absorption spectrum has been obtained
with an angular resolutiion 2.96$\times$2.01 arcsec along PA of 25$^\circ$. The vertical line
indicates the optical systemic velocity of 1850$\pm$20 km s$^{-1}$.
           }
\end{figure}

\section{Summary and concluding remarks}
The results of a radio study of the superwind galaxy NGC1482 which has
a remarkable hourglass-shaped optical emission line outflow as well as
bipolar soft X-ray bubbles of emission, are summarised and discussed
briefly here.

\begin{enumerate}

\item The low-frequency images trace the relatively smooth non-thermal 
emission due to relativistic particles generated largely in the 
starburst. There is no evidence of significant non-thermal 
emission along the optical hourglass-shaped or X-ray outflows.

\item From radio observations we have estimated the supernova rate and
find it to be similar to other well-known starburst galaxies. We also 
estimate the available energy in the
supernovae, which could be the driving force for the superwind outflow. However, 
an estimate of the total energy in all the different components of the outflow
is required before determining whether the energy in the supernovae is adequate
to drive the superwind. 

\item The higher resolution 4860 and 8460 MHz images, which trace the circumnuclear
starburst region,  show a single prominent peak and secondary substructure 
embedded in the more extended emission. These lie at the base of the optical 
hourglass-shaped structure and the X-ray outflows.

\item The prominent radio peak in the high-resolution images has a steep radio 
spectrum between 4860 and 8460 MHz and has not been detected at 14965 MHz. The
highest resolution 8460 MHz A-array image  shows an extension towards the south. 
The brightness temperature of the prominent peak in the 8460 MHz A-array image
is 740$^\circ$ K,  and it is difficult to 
distinguish whether it is an AGN or non-thermal emission from a compact 
starburst region. Optical spectroscopic observations by Kewley et al. (2000)
show no evidence of an AGN.

\item The prominent radio peak is not coincident with the peaks of emission
in the R-band, H$\alpha$ and hard X-ray images, but is coincident with 
the peak in the 2MASS K-band image, suggesting that this feature is possibly 
the centre of the galaxy. The centre of the galaxy estimated from the outer
isophotes of the R-band image is within an arcsec of the radio peak. The geometrical
centre of a line joining the peaks of the red- and blue-shifted 
\hbox{H\,{\sc i}} blobs of emission is within 2 arcsec of the prominent radio peak.

\item  The \hbox{H\,{\sc i}} observations reveal two blobs of emission on
opposite sides of the centre of the galaxy.
The western blob has a maximum red shift of $\sim$190 km s$^{-1}$ relative to
the  systemic velocity of 1850$\pm$20 km s$^{-1}$, while the maximum blue shift for the
eastern one is $\sim$220 km s$^{-1}$.  These blobs are located at 
a distance of $\sim$2 kpc from the centre and have masses of 
approximately 15  and 13.5$\times$10$^7$ M$_\odot$ respectively.

\item  The \hbox{H\,{\sc i}} absorption profile shows multiple components
with the deepest absorption feature at 1836 km s$^{-1}$, consistent with the
systemic velocity of the galaxy. The \hbox{H\,{\sc i}} mass estimated from the
absorption profile is 6.3$\times$10$^6$ M$_\odot$. Relative to the systemic velocity
of 1850$\pm$50 km s$^{-1}$, there is a suggestion of a mild asymmetry in the absorption
profile which could be due to \hbox{H\,{\sc i}} clouds hydrodynamically affected 
by the superwind, in addition to the effects of galactic rotation.

\item Although the distribution of \hbox{H\,{\sc i}} gas is consistent with that seen
for other S0/a galaxies (van Driel 1987; van Driel \& van Woerden 1991), the central
depletion is pronounced, possibly due to the nuclear starburst. 

\end{enumerate}

\section*{Acknowledgments}
We are grateful to Sylvain Veilleux, David Rupke and Dave Strickland for providing us with
FITS files of their optical and X-ray images; Sylvain Veilleux and David Rupke  for 
communicating their results on the systemic velocity of the galaxy prior to publication;
an anonymous referee, and Dave Strickland and Sylvain Veilleux for many valuable and 
critical comments; Jayaram Chengalur, Neeraj Gupta,  Nimisha Kantharia, 
Vasant Kulkarni and Subhashis Roy for their comments on the manuscript. One of us (AH)
thanks the Kanwal Riki scholarship of TIFR for partial financial support.
The GMRT is a national facility operated by the National Centre for Radio 
Astrophysics of the Tata Institute of Fundamental Research. VLA is a operated 
by Associated Universities, Inc. under contract with the National Science 
Foundation. This research has made use of the NASA/IPAC extragalactic database 
(NED) which is operated by the Jet Propulsion Laboratory, Caltech, under 
contract with the National Aeronautics and Space Administration.

\end{document}